\documentclass[]{scrartcl}

\usepackage[utf8]{inputenc}
\usepackage{color}

\usepackage[maxbibnames=99]{biblatex}
\usepackage{booktabs} 

\usepackage[top=2.5cm, bottom=2.5cm, left=2.5cm, right=2.5cm]{geometry}

\usepackage{subcaption}
\usepackage{lscape}
\usepackage{multirow}
\usepackage[normalem]{ulem}
\useunder{\uline}{\ul}{}

\title{Towards a corpus for credibility assessment in software practitioner blog articles}
\subtitle{Preprint}

\author{
    \small{Ashley Williams}\\
\small{Department of Computing and Mathematics} \\
\small{Manchester Metropolitan University, UK} \\
\small{ashley.williams@mmu.ac.uk}
\and
    \small{Matthew Shardlow}\\
\small{Department of Computing and Mathematics} \\
\small{Manchester Metropolitan University, UK} \\
\small{m.shardlow@mmu.ac.uk}
\and
\small{Austen Rainer}\\
\small{School of Electronics, Electrical Engineering and Computer Science} \\
\small{Queens University Belfast, UK} \\
\small{a.rainer@qub.ac.uk}
}

\date{June 2021}

\begin{document}

\maketitle

\begin{abstract}
\textbf{Background}: Blogs are a source of grey literature which are widely adopted by software practitioners for disseminating opinion and experience. Analysing such articles can provide useful insights into the state--of--practice for software engineering research. However, there are challenges in identifying higher quality content from the large quantity of articles available. Credibility assessment can help in identifying quality content, though there is a lack of existing corpora. Credibility is typically measured through a series of conceptual criteria, with 'argumentation' and 'evidence' being two important criteria.

\textbf{Objective}: We create a corpus labelled for argumentation and evidence that can aid the credibility community. The corpus consists of articles from the blog of a single software practitioner and is publicly available.

\textbf{Method}: Three annotators label the corpus with a series of conceptual credibility criteria, reaching an agreement of 0.82 (Fleiss’ Kappa). We present preliminary analysis of the corpus by using it to investigate the identification of claim sentences (one of our ten labels).

\textbf{Results}: We train four systems (Bert, KNN, Decision Tree and SVM) using three feature sets (Bag of Words, Topic Modelling and InferSent), achieving an F1 score of 0.64 using InferSent and a Linear SVM.

\textbf{Conclusions}: Our preliminary results are promising, indicating that the corpus can help future studies in detecting the credibility of grey literature. Future research will investigate the degree to which the sentence level annotations can infer the credibility of the overall document. 
\end{abstract}


\newpage
\tableofcontents
\listoftables

\clearpage

\section{Introduction}
\label{section:introduction}

\subsection{Context and motivation}

Garousi \textit{et al.} \cite{garousi2016need} argue that the wider adoption of grey literature by research would help to bridge the gap between the state--of--art, where research typically operates, and the state--of--practice, what actually happens in industry. Conducting grey literature reviews systematically is problematic, however, as we have limited visibility into how search engines rank results, and the web is a lot larger and more diverse than academic databases. Rainer and Williams \cite{rainer2019using} suggest that the use of credibility assessment as an inclusion criterion in such grey literature reviews can contribute to providing a method for identifying the higher quality results. 

Credibility is, at least in part, a subjective experience. Credibility research attempts to address this subjectivity by reporting credibility assessment concerning a specific user group e.g. the visually impaired \cite{abdolrahmani2016should}, first year students \cite{menchen2011young}, pensioners \cite{liao2010effects}. However, conceptualising credibility at the level of a user group is itself problematic because members of the user group still have unique experiences in time and space. The subjectivity of credibility affects our ability to assess the credibility of web documents, e.g., our ability to consistently annotate web documents. We return to this point in Section \ref{sec:credibility_assessment}.

Credibility is assessed through measuring a series of conceptual criteria. The specific criteria to use depends on the user group and the context of the study being undertaken. Williams and Rainer \cite{williams2019empirical} conducted a survey in order to determine which criteria apply to the credibility assessment of software engineering researchers. Both reasoning and the reporting of experience were considered as important criteria. For measuring reasoning, we can turn to the argumentation mining community for prior research. Measuring experience is more challenging as, according to Rainer \textit{et al.} \cite{rainer2020experience}, there is limited prior research on experience mining.

In this paper, we present a corpus of annotated articles from a single software practitioner's blog (the highly regarded `Joel On Software' blog, written by Joel Spolsky). The articles have been annotated at a sentence level for the presence of argumentation and evidence. The corpus contains 19996 sentences and is publicly available for download. We present results from our preliminary use of the corpus, discuss how the corpus and results build on existing credibility assessment research, contrast the annotations with the results of an existing tool (MARGOT \cite{lippi2016margot}), and consider next steps. 

\subsection{Contributions}
The paper makes the following contributions:

\begin{itemize}
    \item We present a new corpus of 19996 sentences, annotated for argumentation and evidence. The corpus comprises all sentences from 234 blog articles (the full blog roll comprises over 1000 blog articles). The corpus is available for public use\footnote{\url{https://github.com/serenpa/Blog-Credibility-Corpus}}.
    \item To the best of our knowledge, the annotated corpus is the first corpus of blog documents written by an experienced software practitioner and annotated for arguments and evidence.
    \item We also present our preliminary work towards credibility assessment by identifying claim sentences within the corpus.
\end{itemize}


\subsection{Structure of this paper}
The remainder of this paper is structured as follows: section \ref{section:related-work} provides an overview of previous and related work; section \ref{section:dataset-generation} describes the generation of a novel annotated dataset; section \ref{section:method} describes the design of our study; section \ref{section:results} presents and discusses the results of our study; the conclusions and future work are presented in section \ref{section:conclusion}. 
\section{Related work}
\label{section:related-work}

\subsection{Industry--originating research fields}
Grey literature that has been written by experienced software practitioners often influences industry practice. In some cases, ideas originating in industry create and mould new research directions. For example, in 2014, Martin Fowler, a prolific software practitioner, published a blog article on the microservices’ architecture for scalable web applications \cite{lewis_fowler_2014}. According to the blog article, the term “microservices” had been discussed and agreed on by a group of software architects to describe a common architectural style that many of them had been recently exploring. Since the articles' release, microservices have not only become a powerful and widely used architectural pattern in industry (e.g. Netflix famously uses microservices\footnote{\url{https://netflixtechblog.com/tagged/microservices}}, as well as Monzo\footnote{\url{https://www.infoq.com/presentations/monzo-microservices/}}), but also microservices have acquired a large research community (e.g. \cite{soldani2018pains, dragoni2017microservices, zimmermann2017microservices}). Microservices are one example of practice influencing and moulding research.

\subsection{The use of practitioner--generated grey literature in research}
Software engineering researchers often use practitioners as a source of evidence in their studies. Typically, such evidence is collected through traditional evidence gathering techniques, e.g. survey, interview, observation. The world wide web has brought with it a shift in the way that software practitioners disseminate information \cite{storey2014r}. As a result, there is growing interest in utilising grey literature as an additional data source in evidence gathering through Grey Literature Reviews (GLR; e.g., \cite{adams2017shades,adams2016searching,banks2009blog,soldani2018pains}) and Multi--vocal Literature Reviews (MLR; e.g., \cite{garousi2019guidelines}). 

There are however challenges in looking at grey literature. Rainer and Williams \cite{rainer2019using} describe the challenges of looking at blog--like content. These challenges generalise to all grey literature and are summarised here:

\begin{itemize}
    \item Definitions and models -- there exists multiple, and sometimes conflicting definitions for `grey literature.' There are also a lack of models to describe grey literature structure and relationships.
    \item Classification frameworks -- as well as the definitions, there are discrepancies in the literature over how the quality of different grey literature sources compare. Garousi \textit{et al.} \cite{garousi2019guidelines} present a framework for classifying grey literature which builds on existing frameworks \cite{adams2017shades,adams2016searching}. The frameworks imply a hierarchy of quality within different grey literature sources, but Rainer and Williams \cite{rainer2019using} argue that the quality of grey literature cannot be classified solely by the medium in which it is published.
    \item The quantity and quality of grey literature -- the universe of grey literature is substantially larger than academic literature and its quality varies greatly. We need a reliable and rigorous way of filtering the high quality content from the vast quantity available.
    \item Ambiguity of language -- grey literature is typically informal and therefore can use idiomatic structures, which may introduce ambiguity. 
\end{itemize}

\subsection{Credibility assessment}
\label{sec:credibility_assessment}
Assessing the credibility of a document can help distinguish the higher quality grey literature from the vast quantity available. Credibility assessment is subjective to the individual. The literature handles this subjectivity by reporting and assessing conceptual criteria for a particular user group (e.g. visually impaired, first year students, pensioners). Williams and Rainer \cite{williams2019empirical} surveyed software engineering researchers to determine the most important conceptual criteria when assessing blog articles. Two key criteria identified were 1) the presence of the argumentation within the document, and 2) the evidence and personal experience provided to support the argumentation. Personal experience is important in grey literature because where researchers argue based on data and experiment, practitioners form opinions based on their personal and professional experience \cite{devanbu2016belief,rainer2003persuading}. For identifying arguments and experience within text, we can utilise the argumentation mining, opinion mining \cite{salah2019systematic,bakshi2016opinion} and experience mining communities. Lippi and Torroni \cite{lippi2016argumentation} present a review of the state of the argumentation mining community. Lippi and Torroni also released MARGOT \cite{lippi2016margot}, a publicly available tool for assessing a document's argumentation and evidence. MARGOT was trained on the IBM Debater dataset, the largest corpus available at the time. The experience mining community is not as mature as its argumentation mining counterpart. Rainer \textit{et al.} \cite{rainer2020experience} conducted a review of the literature reporting the identification of professional experience in grey literature. The review concluded that more primary studies are needed in order to advance the community. One barrier to such primary studies is a lack of corpora that has been labelled for experience. This paper contributes in that it presents a new dataset which is publicly available and is annotated for argumentation and evidence.

\subsection{Aggregating conceptual criteria measurements into overall credibility}
The subjective nature of credibility also hinders the ability to aggregate conceptual criteria measurements into a score for overall credibility. Previous attempts at ranking using various techniques are easily criticised due to investigators deciding on the weightings of each criteria (e.g. \cite{williams2016identifying}). One solution explored has been to measure and present conceptual criteria (e.g. reasoning and experience) data back to the researcher in tabular format with the ability to rank as they please (such as University ranking tables). Our future research intends to look towards other techniques, such as meta--knowledge \cite{shardlow2018identification}, for ranking and comparing grey literature documents.

\section{Dataset}
\label{section:dataset-generation}

\subsection{The subject of the corpus}
The corpus consists of articles from a single practitioner’s blog, Joel Spolsky\footnote{Spolsky is the co--founder and former CEO of Stack Overflow, and the co--founder of Fog Creek Software (the company that created Trello)}. His blog, `Joel on Software' is widely read and highly regarded by the practitioner community. The blog was mainly active from 2000 to 2012, but still publishes articles sporadically today (the last article published at the time of writing was June 2020). The articles within the blog are a mix of opinion pieces (on subjects such as software and technology, management, and start--ups), advertisements for new products and events, and short casual posts intended for fun, or to provide brief updates to his audience on his thoughts/recent activities. This mix of article types brings with it additional challenges over looking at the blogs of practitioners that maintain a more uniform structure (e.g. Martin Fowler\footnote{\url{https://martinfowler.com/}}).

The blog was chosen due to previous research finding it to be an exemplar of how practitioner--generated content can provide new insights for research. Rainer \cite{rainer2017using} demonstrated the value in analysing practitioner--generated content using a single article from 'Joel on Software,' and Williams \cite{williams2018using} used the blog to evaluate a set of keywords for identifying reasoning.

\subsection{Data gathering}
The data was initially collected using \texttt{COAST\_CRAWL}\footnote{\url{https://github.com/serenpa/coast_crawl}}, a publicly available web crawler. We seed the crawler with the blog's archive page to ensure that all articles are accessible. The crawler initially finds 1693 pages. However, after de--duplication, removing static pages and removing non--article pages, 1023 remain. The article text is extracted using Pattern\footnote{\url{https://github.com/clips/pattern}}.

\subsection{Annotation}
\subsubsection{Tagset}
The articles were annotated for argumentation and evidence. These two criteria were broken down into more specific tags. Dictionary definitions for each of the tags were developed further throughout phase one of the annotations (Section \ref{sec:anno-process}). Table \ref{table:annotation-tags} lists each of the tags with a summary definition for each. 


\begin{table*}[!hbtp]
\small
\centering
\caption{The annotation tags and definitions}
\label{table:annotation-tags}
\begin{tabular}{|p{3cm}|p{10cm}|}
\hline
\textbf{Tag (acronym)}            & \textbf{Definition}                                                                                                                                                                                                                          \\ \hline
\multicolumn{2}{|l|}{\textbf{Argumentation}}                                                                                                                                                                                                                            \\ \hline
Claim (clm)                     & A statement or assertion. This claim may be supported by some reasoning or evidence, and may also be an opinion                                                                                                                             \\ \hline
Reasoning (rsn)                 & Reasons often (but not always) appear close to the claim that the reason is supporting. Reasoning supports a claim with logical justification/explanation                                                                                       \\ \hline
Conclusion (conc)               & A judgement or decision reached by reasoning                                                                                                                                                                                                \\ \hline
\multicolumn{2}{|l|}{\textbf{Evidence}}                                                                                                                                                                                                                                 \\ \hline
Experience (exp)               & References to a personal and/or professional experience which is provided as evidence to support a claim, or reasoning. We are interested here in actual experience (c.f. implied experience or hypothetical experience).         \\ \hline
Event (evnt)                    & Events are defined as things that have happened. Operationally, we may detect events through specific mentions to a moment in time e.g. "Last summer, while attending a conference...". Verbs can also imply that an event has taken place without referencing a specific time e.g. "The boy went to the shops". \\ \hline
Citation (cite)                & May be a URL hyperlink to a other web page, a formal reference in a dedicated references section typically at the end of an article (as in research), a footnote, an in--text citation (without a dedicated references section).              \\ \hline
Code Snippet (code)            & Authors may evidence their claims through the use of code examples. These code examples may be in--text, in a separate block, in an image, or in a table.                                                                                   \\ \hline
Reference to table or image (ref) & Authors may evidence their claims through the use of tables of data and/or images.                                                                                                                                                          \\ \hline
Data/statistic (data)          & Authors may evidence their claims through providing statistics or presenting analysis or other forms of raw/processed data.                                                                                                                 \\ \hline
Other (othr)                    & There may be other forms of evidence which has not been specified in the guidelines. This tag allows for annotators to flag other forms of evidence they may think is relevant for discussion.                                              \\ \hline
\end{tabular}
\end{table*}

\subsubsection{Process}
\label{sec:anno-process}
Annotators were presented with the entire document. They were asked to read the article in its entirety before labelling the sentences which contain argumentation and evidence. WebAnno\footnote{\url{https://webanno.github.io/webanno/}} \cite{eckart-de-castilho-etal-2016-web} was used for annotation. Annotation was conducted in two phases. In phase one, two annotators were employed with an additional third annotator for resolving conflicts. This phase consisted of four rounds of annotation with each of the three annotators completing all of the articles from the round. At the end of each round, we met with the three annotators to discuss annotations as a group. Annotators were encouraged to converge on their annotations, but we allowed discrepancies between the two main annotators which were later resolved by the third annotator. We maintained a set of annotation guidelines which were updated during these meetings as the annotators' definitions evolved and converged. In total, 36 articles were annotated during this initial phase. At the end of the phase, agreement was calculated on the annotations at a sentence level. We used the third annotator to resolve the conflicts between annotators one and two. In instances where a conflict cannot be resolved (e.g., if annotator 3 did not agree with either label), we favour annotator one as they annotated the most articles overall. The agreement at the end of phase one for argumentation labels is 0.817 using Fleiss' Kappa (Po: 0.895; Pe: 0.428), and 0.819 for evidence labels (Po: 0.959; Pe: 0.771).

Phase two then consisted of the annotators working on different articles with no double annotation taking place. In doing so, a further 198 articles were annotated taking the total number of articles up to 234. This leads to a final dataset size of 19996 sentences.

\subsubsection{Description of the dataset}
Table \ref{table:frequency-counts} presents frequency counts and percentages for each of the labels. The table indicates that 34\% of the sentences have no label. The percentages total more than 100\% because approximately 1200 sentences are labelled with more than one label.

\begin{table*}[!hbtp]
\small
\centering
\caption{Frequency counts and percentages for the annotation labels}
\label{table:frequency-counts}
\begin{tabular}{p{4cm}|c|c}
\hline
\textbf{Label} & \textbf{Frequency} & \textbf{Percentage} \\
\hline
Claim &	9202 & 62\\
Reasoning  & 1586 & 11 \\
Conclusion & 331 & 2\\
Citation & 778 & 5\\
Code Snippet & 61 & 0\\
Events & 261 & 2\\
Experience & 2590 & 17\\
Reference to Table or Image & 29 & 0 \\
Statistics or Data & 22 & 0 \\
Other  & 29 & 0  \\
\textbf{Total labelled} & 14889 & \\
No label & 5107 & 34 \\
\hline
\hline
\# documents annotated & 234 & -\\
\# sentences with multiple labels & 1242 & - \\
\hline
\end{tabular}
\end{table*}

Table \ref{table:contingency-table} presents a contingency table of sentences that are labelled with more than one label. Of the 19996 sentences, 1242 sentences have more than one label. We allowed annotators to give more than one label per sentence at their discretion as complex sentences may contain multiple elements to be tagged and therefore do not always align with sentence level annotations.

\begin{table*}[!hbtp]
\small
\centering
\caption{Contingency table (n=1242). The values of zero (0) in the table are included for completeness.}
\label{table:contingency-table}
\begin{tabular}{p{4cm}|c|c|c|c|c|c|c|c|c}
\hline
\textbf{Label} & \textbf{Rsn}	& \textbf{Conc} & \textbf{Cite} & \textbf{Code} & \textbf{Evnt} & \textbf{Exp}	& \textbf{Ref} & \textbf{Data}	& \textbf{Othr}	\\
\hline
Claim	&	643	&	61	&	168	&	44	&	57	&	119	&	2	&	3	&	19		\\
Reasoning	&		&	109	&	19	&	9	&	9	&	53	&	0	&	0	&	2		\\
Conclusion	&		&		&	3	&	1	&	0	&	3	&	0	&	0	&	0		\\
Citation	&		&		&		&	2	&	8	&	88	&	0	&	0	&	3		\\
Code Snippet	&		&		&		&		&	2	&	2	&	0	&	1	&	0		\\
Events	&		&		&		&		&		&	20	&	0	&	1	&	0		\\
Experience	&		&		&		&		&		&		&	0	&	2	&	5		\\
Reference to Table / Image	&		&		&		&		&		&		&		&	0	&	0		\\
Statistics / Data	&		&		&		&		&		&		&		&		&	0		\\
\hline
\end{tabular}
\end{table*}

\subsection{Comparing the annotated dataset with an independent system}

As an exploration of the annotated dataset, we compared the labels assigned by the annotators with labels assigned independently by MARGOT \cite{lippi2016margot}, an established argumentation mining tool. (To clarify: we are not \textit{evaluating} the two set of labels. We view this merely as a comparison to a related system from previous literature). We first used MARGOT to label the sentences of all articles in the Spolsky dataset. MARGOT actually `generated' slightly more sentences (20022 compared to the 19996 sentences  of the annotation dataset). We then selected the 234 articles that had been annotated by both the annotators and MARGOT. 
We matched similar sentences using Jaccard similarity: two sentences were treated as sufficiently similar if the Jaccard similarity score was $>=0.5$. Our matching identified approximately 20000 matched pairs of sentences. For the matched sentences, we compared the labels assigned by the annotators with the labels assigned by MARGOT.

Table \ref{table:confusion-matrices} presents confusion matrices for three sets of labels. In Table \ref{table:confusion-matrices}(a) we compare MARGOT's labelling of a claim (\texttt{TRUE} or \texttt{FALSE}) with the annotators' labelling of \textit{either} a claim, a reason or a conclusion (cf. Table  \ref{table:annotation-tags}). In Table \ref{table:confusion-matrices}(b)  we compare MARGOT's labelling of a claim with the annotators' labelling \textit{only} of a claim. In Table \ref{table:confusion-matrices}(c)  we compare MARGOT's labelling of a claim with the annotators' labelling \textit{only} of either a reason or a conclusion.

\begin{table*}[!hbtp]
\small
\centering
\caption{Confusion matrices comparing the annotators' labels with MARGOT's \cite{lippi2016margot} labels}
\label{table:confusion-matrices}
\begin{tabular}{p{3cm}|p{3cm}|p{3cm}|p{3cm}}
\multicolumn{4}{l}{(a) Contrasting MARGOT's label of a claim  with annotators' labels of a claim,  reason or conclusion.}\\
\multicolumn{4}{c}{}\\
& \multicolumn{2}{c|}{\textbf{MARGOT}} & \\
\textbf{Annotators} & \textbf{\texttt{TRUE}} & \textbf{\texttt{FALSE}} & \textbf{Total}\\
\hline
\textbf{\texttt{TRUE}} &  952  & 9344 & 10296\\
\hline
\textbf{\texttt{FALSE}} & 297 & 9429 & 9726\\
\hline
Total & 1249 & 18773 & 20022\\
\hline
\multicolumn{4}{c}{}\\
\multicolumn{4}{l}{(b) Contrasting MARGOT's label of a claim only with annotators' label of a claim}\\
\multicolumn{4}{c}{}\\
& \multicolumn{2}{c|}{\textbf{MARGOT}} & \\
\textbf{Annotators} & \textbf{\texttt{TRUE}} & \textbf{\texttt{FALSE}} & \textbf{Total}\\
\hline
\textbf{\texttt{TRUE}} & 825 & 8327 & 9152\\
\hline
\textbf{\texttt{FALSE}} & 424 & 10446 & 10870 \\
\hline
Total   & 1249 & 18773 & 20022\\
\hline

\multicolumn{4}{c}{}\\
\multicolumn{4}{l}{(c) Contrasting MARGOT's label of a claim only with annotators' label of a reason or conclusion}\\
\multicolumn{4}{c}{}\\
& \multicolumn{2}{c|}{\textbf{MARGOT}} & \\
\textbf{Annotators}  & \textbf{\texttt{TRUE}} & \textbf{\texttt{FALSE}} & \textbf{Total}\\
\hline
\textbf{\texttt{TRUE}} & 255 & 1542 & 1796\\
\hline
\textbf{\texttt{FALSE}} & 994 & 17220 & 18214 \\
\hline
Total   & 1249 & 18773 & 20022\\
\hline
\end{tabular}
\end{table*}

Table \ref{table:confusion-matrices} shows considerable disagreement between the annotators' labels and MARGOT. We hypothesise that this disagreement can at least partly be explained by definitions and by the nature of the dataset. 

For definitions, we hypothesise that our definition of claim (cf. Table \ref{table:annotation-tags}) is different to the definition applied during the labelling of the IBM Debater dataset that was subsequently used to train MARGOT. For the IBM Debater dataset, Aharoni \textit{et al.} \cite{aharoni2014benchmark} defined a claim as: ``Context Dependent Claim  -- a general concise statement that directly supports or contests the topic''. This contrasts with our definition of a claim as, ``A statement or assertion.  This claim may be supported by some reasoning or evidence, and may also be an opinion.'' (see Table \ref{table:annotation-tags}).

For the nature of the dataset, we hypothesise that the content of the Spolsky dataset is different to that of the IBM Debater dataset. Lippi and Torroni \cite{lippi2016margot} explain that the IBM Debater dataset consisted of 547 Wikipedia articles that had been organized into 58 topics, and annotated with 2294 claims and 4690 evidence facts. By contrast, we have 234 web articles written by one software practitioner on an unknown number of topics. It is likely that the Wikipedia articles will have progressed through more review and revision than the web articles, and will also likely be written in a more formal style. We therefore hypothesise that the annotators' labels in Table \ref{table:confusion-matrices}(c) are most likely to be comparable to the MARGOT claims, however the respective data is too imbalanced to explore this hypothesis further at this stage.






\section{Preliminary Experiment}
\label{section:method}
Our experiments focus on the detection of claims in our dataset. This is only one use of the dataset and the detection of other categories is left to future work (however, preliminary results for our other labels are presented in Table \ref{tab:prelim_results}). The detection of claims is an interesting problem in itself as it is a broadly defined concept and many types of sentences can be considered a claim. In our corpus, 62\% of sentences are annotated as claims. We use 4 approaches as detailed below:

\begin{description}
    \item[BERT:] The BERT masked language model \cite{devlin-etal-2019-bert}, built on the transformer architecture \cite{NIPS2017_7181} is now prevalent at the forefront of NLP research. We used the keras-bert implementation \footnote{\url{https://github.com/CyberZHG/keras-bert}} with the Bert-Large pretrained model (L=24, H=1024, A=16) and configured it with one fully connected hidden layer of 16 units with a ReLU activation function and an output layer of 2 hidden units with a softmax activation function. We used the rectified Adam optimiser \cite{liu2019radam} during training. We report on the model with and without the hidden layer to demonstrate its effect. 
    \item[K-Nearest Neighbour (KNN):] We used the SciKit Learn \cite{scikit-learn} implementation of the KNN \cite{NIPS2004_2566}, with K set to 3. 
    \item[Decision Tree (DT):] We used the SciKit Learn implementation of the decision tree \cite{breiman1984classification}, with a maximum depth of 5.  
    \item[Support Vector Machine (SVM):] We used the SciKit Learn implementation of the SVM \cite{Platt99probabilisticoutputs,chang2011libsvm} with a linear kernel and regularisation parameter $C = 0.25$.
\end{description}

We split the data into train (80\%), validation (10\%) and test (10\%) partitions, which were stratified according to the claim label and shared across all algorithms. We removed any sentences with multiple labels, leaving a total of 8539 Claim sentences and 9612 sentences with no label (split evenly across the three partitions). The validation partition was used to measure the loss whilst training BERT and to select appropriate algorithms from Sci-kit learn. The final results are reported on the test partition. We trained Bert for one epoch in each case and did not further tune the hyperparameters of BERT or the other algorithms. Although this can lead to a perceived improvement in results, it also often leads to model overfitting, which we sought to avoid.

We created three feature sets that were used as input to our machine learning algorithms (excluding BERT, which does not require external features). These are described as follows:

\begin{description}
    \item[Bag of Words:] In this approach we first computed the mutual information \cite{novovivcova2004feature} between each word and the class label. We selected the top 100 words as binary features, which indicated the presence of a word that distinguished the class. 
    \item[Topic Modelling:] We first created a document-token matrix indicating the frequency of each token in each sentence. We then used the gensim \cite{rehurek_lrec} implementation of LDA \cite{blei2003latent} to reduce the dimensionality of the matrix and to create topic vectors for each document in a technique commonly known as topic modelling. This technique forces words which occur in similar contexts to be in the same topic. We limited the vector size to 100 topics.
    \item[InferSent:] We used the Facebook library InferSent \cite{conneau-EtAl:2017:EMNLP2017}, which provides a 4096 dimensional embedding for a given sentence. InferSent uses FastText vectors \cite{grave2018learning} to get the embedding for each token and then passes these through a pre--trained model which identifies the importance and weighting of each embedding before recombination. Each dimension of the resulting embedding was used as a feature to the algorithm, giving 4096 distinct features.
\end{description}

We ran each algorithm with each feature set and report the results in Table \ref{tab:results}. We attempted to combine feature sets, but this did not lead to any improvement in the overall scores and so these results are omitted.
\section{Results \& Discussion}
\label{section:results}

\begin{table*}[ht]
    \centering
    \begin{tabular}{c c|c c c}
        Classifier & Features & Precision & Recall & F1  \\\hline
        \multirow{2}{*}{BERT}     & No Hidden Layer & 0.54 & 0.48 & 0.51 \\
             & Hidden Layer    & 0.75 & 0.48 & 0.58 \\\hline
             & BOW             & 0.88 & 0.12 & 0.20 \\
        KNN  & Topic Modelling & 0.52 & 0.45 & 0.48 \\
             & InferSent       & 0.57 & 0.58 & 0.58\\\hline
             & BOW              & 0.82 & 0.02 & 0.04 \\
        DT   & Topic Modelling & 0.56 & 0.31 & 0.40 \\
             & InferSent       & 0.57 & 0.66 & 0.61\\\hline
             & BOW             & 0.85 & 0.01 & 0.02 \\
        SVM  & Topic Modelling & 0.00 & 0.00 & 0.00 \\
             & InferSent       & 0.62 & 0.65 & \textbf{0.64}\\
    \end{tabular}
    \caption{The results of detecting claim sentences}
    \label{tab:results}
\end{table*}

Our results are shown in Table~\ref{tab:results}. We tried using Bert as it is has been shown to achieve state--of--the--art performance with little fine tuning for other NLP tasks. The results however, were somewhat disappointing. Even the addition of a hidden layer did not greatly improve the scores. Although we could have spent time further modifying the network structure and training for many epochs, we instead decided to use classic machine learning algorithms from sci--kit learn. 

We used 2 traditional feature generation techniques (Bag of Words and Topic Modelling) and one state of the art method of generating sentence embeddings (InferSent). The InferSent features outperformed Bag of Words and Topic Modelling with every classifier. This is surprising as Bag of Words and Topic Modelling have 100 features each, compared to 4096 features for InferSent. Typically, the performance of classical machine learning algorithms decreases when presented with many features. We hypothesise that many of the dimensions in the InferSent embeddings were not being used as part of the classification strategy. Our best performing system used the InferSent embeddings and a Linear SVM and received an F1 score of 0.64, with precision at 0.62 and recall at 0.65. This indicates that the claim vs. non-claim sentences are separable and that a model can be built to distinguish between them. InferSent provides an embedding for a sentence and it would be interesting to analyse which parts of a sentence are being used in the classification of claims vs. non--claims.

In one instance, our model was not able to produce a reliable model for the Claim sub--dataset (see SVM + Topic Modelling). We used a linear SVM and this implies that no linear boundary could be found in the topic space to separate our classes. Note that the KNN and Decision Tree, both of which can create complex boundaries in feature space, were able to produce models using the topic model features. It may be the case that using a kernel--based SVM would yield a more reliable model for the Topic--Modelling features, however we avoided non--linear SVM as the time taken to train is too great with our high--dimensional InferSent embeddings.

It may well be possible to improve our scores by tuning the algorithms used or using a more powerful Transformer based architecture, however we have focused our results on building a model for claims as a benchmark for our dataset. We expect to further improve on the scores we have reported as well as identify other elements in our corpus in future work.


More generally, credibility assessment is a difficult task as it is subjective, multi--disciplinary, and lacking in formal definitions. The credibility literature acknowledges that argumentation and the reporting of evidence are important criteria. However, when we try to apply them to this particular application, we find a relatively low amount of reasoning and evidence within the dataset. Furthermore, most studies within the argumentation mining community focus on well--structured text such as legal texts, persuasive essays and debate corpora. When applying technologies such as MARGOT, to web documents, the classification task becomes more difficult as the writing style is more casual, less structured and at least sometimes more ambiguous than one might expect of better structured texts.

Finally, the credibility assessment community recognises a need for formal definitions for both the conceptual criteria analysed and the term `credibility' itself \cite{williams2017analysis}. We see similar issues when contrasting our work with the argumentation mining community. For example, the IBM Debater dataset uses the term `claim' as its measure of argumentation. In this paper, we define claim as synonymous with assertion and opinion, distinguishing it from reasoning and conclusion.

\section{Conclusions \& Future Research}
\label{section:conclusion}

\subsection{Threats to Validity}
There are multiple threats to this research, with each threat also providing an opportunity for future work:

\begin{enumerate}
    \item The metrics used for assessing credibility are based on the findings of a previous literature review and survey \cite{williams2019empirical}. The literature review was not conducted systematically and only thirteen papers were selected for analysis. A broader, systematic review may yield new important metrics for credibility assessment. In addition, although the response rate of the survey was good, the overall number of responses in comparison to the community of software engineering researchers is relatively low. Further research is needed to ensure that the credibility metrics do not only apply to this subset of researchers.
    \item There are threats with the way in which we have conducted our annotations. Our student annotators may have different definitions for our credibility metrics than software engineering researchers (our target demographic). This may in turn lead to different annotations. Future research will look at the quality of our annotations.
    \item In looking at only one source of grey literature, a single practitioners blog, it is unclear how well our models, and models trained using our dataset, generalise to other sources. Future research will investigate the degree to which our metrics and models work over other data sources (e.g. Twitter, Stack Overflow, GitHub).
\end{enumerate}

\subsection{Future research}
There are many avenues open for future research. This paper presents preliminary analysis of one label within the dataset, further analysis across all labels is a natural next step. The analysis could then be aggregated and ranked to form an overall credibility rating for each individual article. The credibility ratings and quality could then be compared against one another. 

We also plan to benchmark against other tools and to look at particular subsets of the dataset. For example, we plan to look more closely at sentences where MARGOT and the annotators agree. This is challenging as the result would be a small, imbalanced dataset, however it could provide further insight into identifying quality content.

Finally, the dataset presented in this paper can also aid further work in each sub--community (e.g. argumentation, experience, opinion mining). For example, previous work \cite{williams2018software,williams2019software} has investigated the degree to which practitioners cite research in their blog articles. The dataset allows for further, more in--depth analysis of citations. Similarly, another area for future research is the dataset's potential to be used as a source in future experience mining primary studies.

\subsection{Conclusions}
In this paper we have presented a new dataset annotated for elements of argumentation and evidence. The dataset comprises 19996 labelled sentences from 234 complete blog articles, with all articles written by an experienced software practitioner. Our intention is that the dataset can help future studies in automating credibility assessment and the comparison of documents for ranking based on credibility and quality. The dataset is publicly available\footnote{\url{https://github.com/serenpa/Blog-Credibility-Corpus}}.

In addition to the dataset generation, we present preliminary analysis toward automating the identification of claim sentences, one of our ten labels. An SVM trained using the InferSent feature set provides a F1 score of 0.64.

\begin{landscape}
\begin{small}

\begin{table*}[ht]
\centering
\small
\begin{tabular}{p{2.5cm}|p{1.6cm}|p{0.9cm}|p{0.8cm} p{0.8cm} p{0.8cm} p{0.8cm}|p{0.8cm} p{0.8cm} p{0.8cm} p{0.8cm}|p{0.8cm} p{0.8cm} p{0.8cm} p{0.8cm}}

                                                            &                              &                         & \multicolumn{4}{c|}{\textbf{TF/IDF}}              & \multicolumn{4}{c|}{\textbf{Topic Modelling}} & \multicolumn{4}{c}{\textbf{BERT}}                \\ \hline
\textbf{Label Set}                                          & \textbf{Balanced Class Size} & \textbf{Total Data}     & P     & R     & F1                        & Acc   & P         & R         & F1        & Acc       & P     & R     & F1                        & Acc   \\ \hline
Reasoning +  Conclusion                                     & 1808                         & 3616                    & 0.609 & 0.588 & 0.598                     & 0.622 & 0.624     & 0.683     & \textbf{0.652}     & 0.651     & 0.593 & 0.654 & 0.622                     & 0.619 \\ \hline
Reasoning + Conclusion + Claim                              & 9646                         & 19292                   & 0.655 & 0.726 & \textbf{0.689}                     & 0.67  & 0.591     & 0.64      & 0.614     & 0.596     & 0.597 & 0.696 & 0.642                     & 0.611 \\ \hline
Reasoning                                                   & 1586                         & 3172                    & 0.601 & 0.619 & 0.61                      & 0.639 & 0.547     & 0.63      & 0.585     & 0.594     & 0.607 & 0.685 & \textbf{0.644}                     & 0.655 \\ \hline
Conclusion                                                  & 331                          & 662                     & 0.625 & 0.583 & \textbf{0.603}                     & 0.654 & 0.471     & 0.533     & 0.5       & 0.519     & 0.469 & 0.633 & 0.539                     & 0.511 \\ \hline
Claim                                                       & 9202                         & 18404                   & 0.634 & 0.698 & \textbf{0.664}                     & 0.644 & 0.578     & 0.632     & 0.604     & 0.582     & 0.593 & 0.655 & 0.622                     & 0.6   \\ \hline
Evidence   (Citation/ code/events/exp erience/ref/ stats/Other) & 3636                         & 7272                    & 0.665 & 0.597 & \textbf{0.629}                     & 0.661 & 0.551     & 0.636     & 0.591     & 0.576     & 0.58  & 0.587 & 0.583                     & 0.597 \\ \hline
Code Snippet                                                &  61       & 122 & 0.75  & 0.6   & 0.667                     & 0.76  & 0.75      & 0.3       & 0.429     & 0.68      & 0.727 & 0.8   & \textbf{0.762} & 0.8   \\ \hline
Events                                                      & 261                          & 522                     & 0.718 & 0.571 & 0.636                     & 0.695 & 0.531     & 0.531     & 0.531     & 0.562     & 0.656 & 0.816 & \textbf{0.727}                     & 0.714 \\ \hline
Experience                                                  & 2590                         & 5180                    & 0.711 & 0.638 & \textbf{0.672} & 0.69  & 0.612     & 0.63      & 0.621     & 0.617     & 0.617 & 0.605 & 0.611                     & 0.616 \\ \hline
Reference to Table/Image                                    & 29       & 58  & 1     & 0.333 & 0.5                       & 0.667 & 0         & 0         & 0         & 0.25      & 0.571 & 0.667 & \textbf{0.615}                     & 0.583 \\ \hline
Statistics/Data                                             & 22       & 44  & 1     & 0.333 & 0.5                       & 0.556 & 0         & 0         & 0         & 0.333     & 1     & 0.833 &\textbf{ 0.909} & 0.889 \\ \hline
Other                                                       & 29       & 58  & 0.75  & 0.5   & 0.6                       & 0.667 & 0         & 0         & 0         & 0.333     & 0.8   & 0.667 & \textbf{0.727}                     & 0.75  \\ \hline
Citation                                                    & 778                          & 1556                    & 0.597 & 0.537 & 0.565                     & 0.606 & 0.481     & 0.517     & 0.498     & 0.503     & 0.522 & 0.557 & \textbf{0.539}                     & 0.545 \\ \hline
\end{tabular}
\caption{Preliminary results of detecting other labels within the dataset using Random Forests}
\label{tab:prelim_results}
\end{table*}
\end{small}
\end{landscape}

\section*{Acknowledgements}
We thank the annotators who contributed to this project, the Centre for Advanced Computational Science for their seed funding, and the reviewers for their valuable feedback.

We have sought advice on the ethics of analysing blogs. All of the blog articles that we have used are publicly available and we would like to thank Joel Spolsky, who is aware of this research.

\printbibliography

\end{document}